\documentclass[a4paper,english]{article}
\usepackage[T1]{fontenc}
\usepackage{graphicx}
\usepackage[authoryear]{natbib}

\usepackage{graphics}
\bibpunct{(}{)}{;}{a}{}{,}
\usepackage{times}

\usepackage{babel}

\begin{document}

\title{Radial velocities of {}``slow movers'' -- call for observations}

\author{Piotr A. Dybczy\'{n}ski$^{1}$, Tomasz Kwiatkowski$^{2}$ \\
\\
$^{1,2}$Astronomical Observatory of the A.~Mickiewicz University,\\
S\l{}oneczna~36, 60-286~Pozna\'{n}, Poland, \\
e-mails: $^{1}$dybol@amu.edu.pl,
$^{2}$tkastr@amu.edu.pl}

\date{Received ..... / Accepted......}
\maketitle
\begin{abstract}
This paper presents a list of suggested stars for radial velocity
measurements. We explain here in brief the research project for which
the radial velocity of the {}``slow movers'' i.e. small proper motion
stars are necessary. Basing on this study we prepared a list of 1100
stellar targets with very accurate positions, proper motions and trigonometric
parallaxes but without radial velocity measurements. Distributions
of stellar brightnesses and spectral types among these stars are presented
as well as its {}``most wanted'' subset. We announce the begin of
the radial velocity measurements to be conducted in our observatory
and offer some coordination for observations of targets that cannot
be reached from our location.

Keywords: Stars: kinematics -- Techniques: radial velocities 
\end{abstract}

\section{Introduction}

During our research on the long-period comets origins we simulate
the mechanisms of production of the observable comets directly form
the Oort\citeyearpar{oort:1950} cometary cloud by means of stellar
and galactic perturbations. \textbf{}A natural consequence of that
study would be \textbf{}a search for stellar perturbers which passed
near the Sun during the last several million years or will have such
a chance in the future. Such passages can induce strong asymmetries
in the observed population of the long-period comets, as it was showed
by \citet{dyb-teneryfa:2002,dyb-asymm:2002,dyb-hab1:2002}. \textbf{}Additionally,
such passages should be taken into account when studying the past
motion of observed long-period comets as it was demonstrated in \textbf{\citet{dyb-hist:2001}.}

In our first attempt to complete a list of potential stellar perturbers
\citep{dyb-kan:1999} we found, that the majority of stars with measured
parallax (most of them are from the Hipparcos catalogue, \citealp{HIPPARCOS})
do not have their radial velocities measured what excludes them from
further consideration. Similar conclusions were reported by \citet{garcia-sanchez:1999,garcia-sanchez:2001}:
they could attribute radial velocity values to less than 50\% of the
selected candidate stars. The lack of radial velocities in such cases
is most often a consequence of the selection criteria used by observers\textbf{.}
When searching for the promising candidate for radial velocity measurements
they prefer stars which high proper motion suggests high space velocity.
\textbf{}Thus {}``fast movers'' (in the sense of high proper motion)
are highly favored. In our research we are interested in the stars
moving almost directly towards (or outwards) the Sun what implies
very small proper motion. We will call these stars {}``slow movers''.

\section{\label{sec:Missing-radial-velocities}Missing radial velocities}

In our study on the stellar perturbations of the Oort cloud comets
we decided to use the most precise subset of the Hipparcos catalogue
\citep{HIPPARCOS}, the ARIHIP catalogue \citep{ARIHIP}, containing
the results of the Hipparcos mission re-reduced for better proper
motion calculation\textbf{s} with the aid of the precise ground based
astrometric catalogues\emph{.} ARIHIP is available on the Internet
at \emph{http://www.ari.uni-heidelberg.de/arihip/}. For our purpose
we used the {}``long term'' solutions included in ARIHIP because
they seem to be the best available approximations of the long term
motion of these stars. For a more detailed explanation see \citet{wielen:2001}.

Our procedure of selecting candidate stars is very similar to that
used by \citet{garcia-sanchez:1999,garcia-sanchez:2001}. For all
stars with the reliable parallax in ARIHIP we calculated the miss-distance
between a star and the Sun basing on the straight line motion model.
As it was shown this simple model is quite sufficient as the first
selection tool. In our first list of candidates (hereafter LC) we
included all stars with the calculated miss-distance less than 5 pc
assuming for all stars $v_{r}=100$ km\,s$^{-1}$. This list consisted
of 2123 stars. However, there are radial velocities included in ARIHIP
for some stars so in the second step we used these values and again
adopted $v_{r}=100$ km\,s$^{-1}$ for the rest of the stars. We
found ARIHIP velocities for 824 stars and as a consequence 738 of
them have been rejected because of the recalculated miss-distance
greater than 5~pc. Only 86 were adopted as the Oort cloud perturbers
(hereafter LP). 

Next we searched other sources for the missing radial velocities.
The most helpful were the ADS (\emph{http://adswww.harvard.edu}) and
CDS (\emph{http://cdsweb.u-strasbg.fr}) services as well as the Bibliographic
Catalogue of Stellar Radial Velocities available on the Internet (\emph{http://www.casleo.gov.ar/catalogo/catalogoin.htm}).
The latter is the 2002 update of the previously published version
\citep{malaroda:2001}. As a result we found additional radial velocity
values for 199 stars, 181 of which were subsequently excluded from
the list: after recalculations their miss-distance become greater
than 5 pc. The remaining 18 stars were added to the list of stellar
perturbers (LP). 

Finally we obtained two lists: LP, with the confirmed stellar perturbers
among the ARIHIP stars with miss-distance less than 5 pc and the LC
list of potential perturbers which radial velocities are still unknown\textbf{.}
A detailed description of our LC list and its {}``most wanted''
subset are presented in Sect.\ref{sec:Most-wanted-target} while the
influence of the stars from the LP list on the Oort cloud comets motion
will be discussed in detail in the separate paper, \emph{}in preparation.

\begin{figure}
\includegraphics[%
  height=1.0\columnwidth,
  keepaspectratio,
  angle=270]{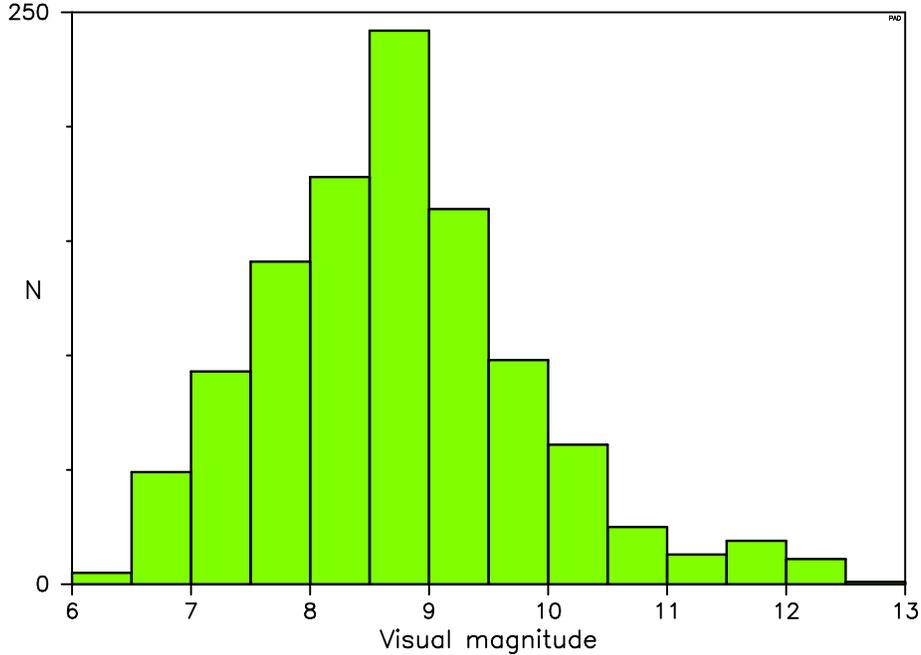}

\caption{\label{dist-magnitude}The distribution of the visual magnitude (copied
from the ARIHIP catalogue) for the proposed 1100 target stars. The
significant decrease of number of stars fainter than 9 mag is due
to the Hipparcos catalogue limit.}
\end{figure}

\section{Observing plans for slow movers}

As the number of slow movers is quite large, reliable measurements
of their radial velocities will deserve a significant amount of observing
time. The accuracy needed to filter out most of the non-significant
objects (with relatively low $v_{r}$ ) is not particularly high and
thus it does not put a special demand on the instrumentation to be
used. When found, slow movers with a high radial velocity can be studied
further with a better spectrograph, yielding absolute accuracy on
the level of $0.1$~km\,s$^{-1}$.

\begin{figure}
\includegraphics[%
  height=1.0\columnwidth,
  keepaspectratio,
  angle=270]{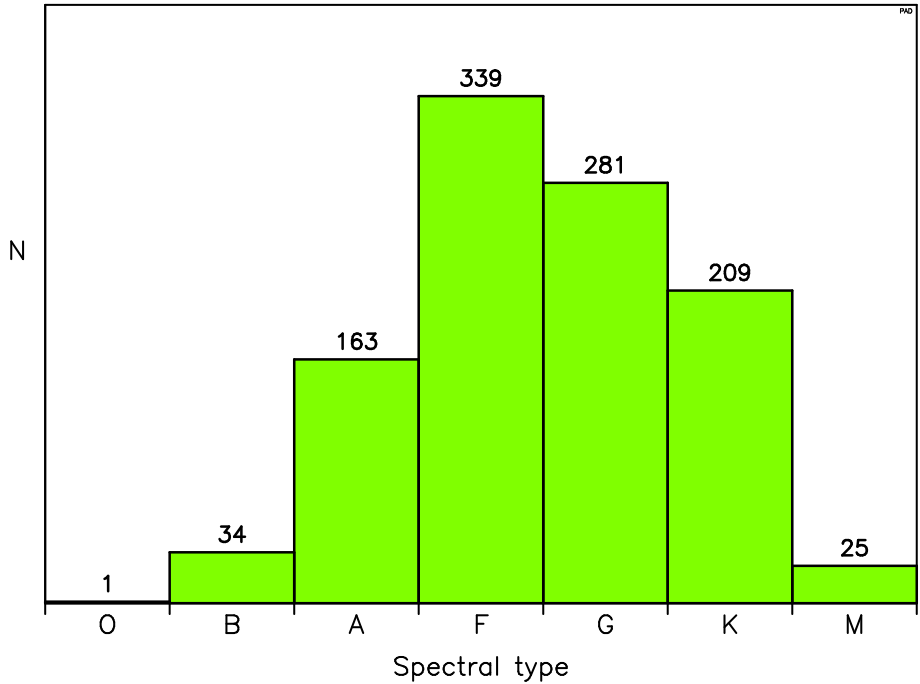}

\caption{\label{dist-spectra}The distribution of the spectral type among
the proposed targets. The number of stars in each type are shown on
top of the corresponding bar. For 1100 stars present in our list we
found 1052 spectral type descriptions, mainly in the Tycho-2 Spectral
Type Catalogue \citep{wright:2003}. It should be noted that some
of the spectral type assignments are very approximate. }
\end{figure}

There were several projects aiming at measurements of radial velocities
of Hipparcos stars. Up till now they obtained radial velocities of
several thousands objects (see for example \citet{grenier:1999,nidever:2002}),
but their selection criteria were different from ours so it may take
a lot of time before all slow movers are measured.

To speed up the process of building a database of slow movers radial
velocities we would like to get observers interested in this group
of star. To avoid unnecessary duplication of efforts we have set up
the LC list of candidates for perturbers which is available on the
web. The immediate aim of this would to be to mark all stars that
have already been measured even if the results are not publicly available
yet. Later, as new radial velocities are being published, the list
will be gradually changed into a database, which can easily be used
for different research projects.

We are also planning our own observations with a newly build fiber-fed
echelle spectrograph. While the main aim of this instrument will be
radial velocity curves of binary stars, slow movers will also be measured
during its free time.

\begin{table*}

\caption{\label{tab-most-wanted}The {}``most wanted'' list: 13 stars which
have the Sun miss-distance less than 3 pc for the assumed radial velocity
$v_{r}=20$ km\,s$^{-1}$ only.}

\begin{center}\begin{tabular}{|r|r|c|c|r|r|c|}
\hline 
\multicolumn{1}{|c|}{HIP}&
\multicolumn{1}{c|}{TYC}&
DD&
V{[}mag{]}&
\multicolumn{1}{c|}{$\alpha$}&
\multicolumn{1}{c|}{$\delta$}&
spectal type\tabularnewline
\hline
\hline 
6935&
8036 01074 1&
2.432&
8.90&
01 29 23.2&
-47 56 23&
F5 IV/V\tabularnewline
\hline 
15929&
4715 00907 1&
1.444&
8.44&
03 25 10.6&
-06 44 08&
F5 V\tabularnewline
\hline 
19527&
0074 00247 1&
1.605&
 8.50&
04 11 00.5&
+01 57 52&
A9 V\tabularnewline
\hline 
24124&
0111 00834 1&
0.313&
10.94&
05 10 52.4&
+06 16 28&
\tabularnewline
\hline 
24600&
4764 00700 1&
1.495&
9.58&
05 16 36.5&
-06 35 21&
A2 IV/V\tabularnewline
\hline 
25469&
0105 00438 1&
2.643&
8.51&
05 26 48.1&
+02 04 06&
 B8 V\tabularnewline
\hline 
29035&
3386 00348 1&
1.869&
8.92&
06 07 32.1&
+51 57 32&
G5\tabularnewline
\hline 
30108&
3375 00998 1&
0.823&
8.03&
06 20 09.5&
+46 38 49&
G5\tabularnewline
\hline 
30344&
6510 01219 1&
1.335&
7.37&
06 22 57.7&
-24 33 22&
K0/1 V\tabularnewline
\hline 
38205&
9192 01237 1&
2.844&
8.91&
07 49 39.1&
-74 38 15&
F6 V\tabularnewline
\hline 
56798&
5513 00809 1&
1.356&
8.73&
11 38 38.8&
-11 14 20&
 G3 V\tabularnewline
\hline 
84263&
9052 01617 1&
1.778&
8.50&
17 13 30.2&
-60 30 40&
F7 V\tabularnewline
\hline 
112584&
4477 01185 1&
2.761&
9.12&
22 48 07.6&
+69 04 30&
G0\tabularnewline
\hline
\end{tabular}\end{center}
\end{table*}

Our new instrument allows us to measure radial velocities of late
spectral type stars down to 8-9~mag, with 30-60~min exposures. The
location of the telescope ($\phi=52^{\circ}$) sets additional constraints
on the available targets. Going through our slow movers list and selecting
stars with $\delta>-8^{\circ}$ and spectral type later then F0 we
obtain 18 stars with visual magnitude $V<7$~mag, 94 stars with $7<V<8$~mag
and 158 stars with $8<V<9$~mag. We plan to observe them starting
from the brightest end and moving on to the dimmer ones. As we progress,
our list available at the web-page will be updated.

\section{\label{sec:Most-wanted-target}Proposed targets}

As it was described in Sec.\ref{sec:Missing-radial-velocities} after
using all radial velocity values available to us we shortened the
LC list to 1100 stars. All of them have the Sun miss-distance less
than 5~pc assuming $v_{r}=100$ km\,s$^{-1}$ . In Figs. \ref{dist-magnitude}
and \ref{dist-spectra} we present distributions of the visual magnitude
and spectral type among proposed targets. As it can be easily noted,
these stars are relatively bright and the majority of them are of
late spectral types. The proposed targets list (LC) is available on
the web at the address: \emph{http://www.astro.amu.edu.pl/Science/SlowMovers}/. 

The main list is named \emph{targets.lst.} The first column gives
the Hipparcos number of the star (strictly ascending), next is the
Tycho-2 corresponding number and the third column shows the calculated
Sun miss-distance ($DD$) in parsecs. The lower is the $DD$ value
the higher is our interest in this particular object, ie. the probability
of confirmation that the star is actually the Oort cloud perturber.

For the user convenience in the next four columns we added the visual
magnitude of the star, its position in the sky (J2000 Right Ascension
and Declination) and its spectral type copied from the Tycho-2 Spectral
Type Catalog \citep{wright:2003}, recently made available at CDS.
>From our 1100 stars 96 are not included in this catalogue but with
the help of SIMBAD database we decreased the number of missing spectra
down to 48 stars.

The calculated Sun miss-distance of each star on the LC list is computed
for the assumed $v_{r}=100$ km\,s$^{-1}$ . To obtain the {}``most
wanted'' group of targets we repeated all the calculations, now assuming
$v_{r}=20$ km\,s$^{-1}$ and limiting the Sun miss-distance $DD$
to be smaller than 3~pc. This gave us a list of 13 stars of particularly
high probability of confirmation as an Oort cloud perturber. They
are listed in Tab.\ref{tab-most-wanted}.

\section{Conclusions}

As it was mentioned above, the main purpose of this paper is to raise
the interest in measuring radial velocities of the proposed 1100 low
proper motion stars. Due to the current location of our instrument
we cannot observe stars having declinations below $-8^{\circ}$. There
are 572 such stars in our list. We are also restricted to the brighter
end of the list but still the number of stars is quite large: it takes
time to determine the absolute radial velocity of the star, especially
when several subsequently obtained values are not consistent with
each other.

So, here is the {}``call for cooperation''. We offer some kind of
coordination by marking as {}``done'' already measured stars or
marking as {}``planned'' other targets upon the information obtained
from observers. First such marks were added to our list during the
preparation of this manuscript. Following the private communications
from Dave Latham we marked stars that have several unpublished radial
velocity measurements made with CfA Digital Speedometers (238 such
stars) and stars included in the CfA list of 10000 main-sequence stars
closer than 100 pc and that have at least one new CfA radial velocity
(unpublished) measurement (142 such stars). 

\textbf{Acknowledgements}.This research has made an extensive use
of the SIMBAD database, operated at CDS, Strasbourg, France and has
made use of NASA's Astrophysics Data System. This work was partially
supported by the KBN Grants no. 2P03D01324 (PAD) and 5P03D00220 (TK).
This manuscript was prepared with the LyX \TeX front-end under
the Debian GNU Linux.

\bibliographystyle{apj}
\bibliography{comets3}

\end{document}